\documentclass[graybox]{svmult}

\usepackage{mathptmx}       % selects Times Roman as basic font
\usepackage{helvet}         % selects Helvetica as sans-serif font
\usepackage{courier}        % selects Courier as typewriter font
\usepackage{type1cm}        % activate if the above 3 fonts are
\usepackage{makeidx}         % allows index generation
\usepackage{graphicx}        % standard LaTeX graphics tool
\usepackage{multicol}        % used for the two-column index
\usepackage[bottom]{footmisc}% places footnotes at page bottom

\makeindex             % used for the subject index
\begin{document}

\title*{ Transmission of packets on a hierarchical network: Avalanches, statistics and explosive percolation}
\titlerunning{Transmission of packets on a hierarchical network}
\author{\bf Neelima Gupte and Ajay Deep Kachhvah}
\institute{Neelima Gupte \at Department of Physics, Indian Institute of Technology, Madras, Chennai, 600036, India. \email{gupte@physics.iitm.ac.in}
\and Ajay Deep Kachhvah \at Nonlinear Physics Division, Institute for Plasma Research, Bhat, Gandhinagar 382428, Gujarat, India. \email{ajaydeep@ipr.res.in}}
\maketitle

\abstract*{We discuss transport on load bearing branching hierarchical networks which can model diverse systems which can serve as models of river networks, computer networks, respiratory networks and granular media. We study avalanche transmissions and directed percolation on these networks, and on the V lattice, i.e., the strongest realization of the lattice. We find that typical realizations of the  lattice show multimodal distributions for the avalanche transmissions, and a second order transition for directed percolation. On the other hand, the V lattice shows power-law behavior for avalanche transmissions, and a first order (explosive) transition to percolation. The V lattice is thus the critical case of hierarchical networks. We note that small perturbations to the V lattice destroy the power-law behavior of the distributions, and the first order nature of the percolation. We discuss the implications of our results.}

\abstract{We discuss transport on load bearing branching hierarchical networks which can model diverse systems which can serve as models of river networks, computer networks, respiratory networks and granular media. We study avalanche transmissions and directed percolation on these networks, and on the V lattice, i.e., the strongest realization of the lattice. We find that typical realizations of the  lattice show multimodal distributions for the avalanche transmissions, and a second order transition for directed percolation. On the other hand, the V lattice shows power-law behavior for avalanche transmissions, and a first order (explosive) transition to percolation. The V lattice is thus the critical case of hierarchical networks. We note that small perturbations to the V lattice destroy the power-law behavior of the distributions, and the first order nature of the percolation. We discuss the implications of our results.}

\section{Introduction}
\label{sec:1}

The study of transport processes on networks of varying types has attracted much recent interest \cite{duch, holme}, and is important from the point of view of applications. Earlier studies of transport processes on networks has been carried out for important classes like scale free and random networks \cite{tadic, ohira, zhao}. However, an important class of networks, viz., that of branching hierarchical networks \cite{pre11, pre12}, has not yet been extensively explored. Networks of the branching, hierarchical type are common in both real and engineering contexts. Examples of such networks include river networks \cite{river}, models of granular media \cite{copper}, and voter models \cite{voter,liggett}, the Domany Kinzel cellular-automata model \cite{domany} and the branching hierarchical model of the lung inflation process \cite{suki, suki2}. In this paper, we explore transport processes such as avalanche dynamics, and percolation on typical branching hierarchical lattices, and on the V lattice, a special realization of such lattices. We show that the behavior of transport on the V lattice is very distinct from that on typical realizations, and the V lattice is thus the critical realization of the branching hierarchical lattices. We also show that this special behavior is soon lost when the lattice is perturbed, indicating its critical nature. We also discuss the implications of our results.

\section{The Model}
\label{sec:2}
The base model for the load bearing hierarchical network \cite{pre11, janaki} discussed here  is a regular $2D$ lattice of sites, where each site is connected at random to exactly one of its two neighbors in the layer below. The choice of the connection between the left and right neighbors to a given site $i^D$ at any $D$th layer is made with some probability $p$, where ($0<p<1$), for a connection to the left neighbor ${\it i_l^{D-1}}$, and probability $(1-p)$ to the right neighbor ${\it i^{D-1}_r}$. 
Each site has the capacity to bear unit weight if it is not connected to either of its neighbors in the layer above, and can bear weight $w+1$ if it is connected to site(s) whose capacities add up to $w$, in the layer above. Thus, the capacity $w(i^D)$ of the $i^{th}$ site in the $D^{th}$ layer is given by the equation
\begin{equation}
w(i^D)=l(i^{D-1}_l,i^D)w(i^{D-1}_l)+l(i^{D-1}_r,i^D)w(i^{D-1}_r)+1
\end{equation}
The quantity $l(i^{D-1}_l,i^D)=1$ if a connection exists between ${\it i^{D-1}_l}$ and ${\it i^{D}}$, and $0$ if otherwise. The network consists of clusters, where a cluster is a set of sites connected with each other. The trunk is defined as the set of connected sites in the largest cluster with the highest weight bearing capacity in each layer. The sum of the weight bearing capacities of the sites along the trunk is defined as the trunk capacity $W_T$ of the given realization of the network. This model has similarity with the critical case $q(0,1)$ of model of granular media \cite{copper}, models of river networks \cite{river}, and the Takayasu model of the aggregation process with injection \cite{takayasu}. Other models  analogous to our model are voter models \cite{voter,liggett}, the Domany-Kinzel cellular-automata model \cite{domany}, and the branching hierarchical model of the lung inflation process \cite{suki, suki2}.

\begin{figure}[htb]
\begin{center}
\includegraphics[height=5.5cm,width=10cm]{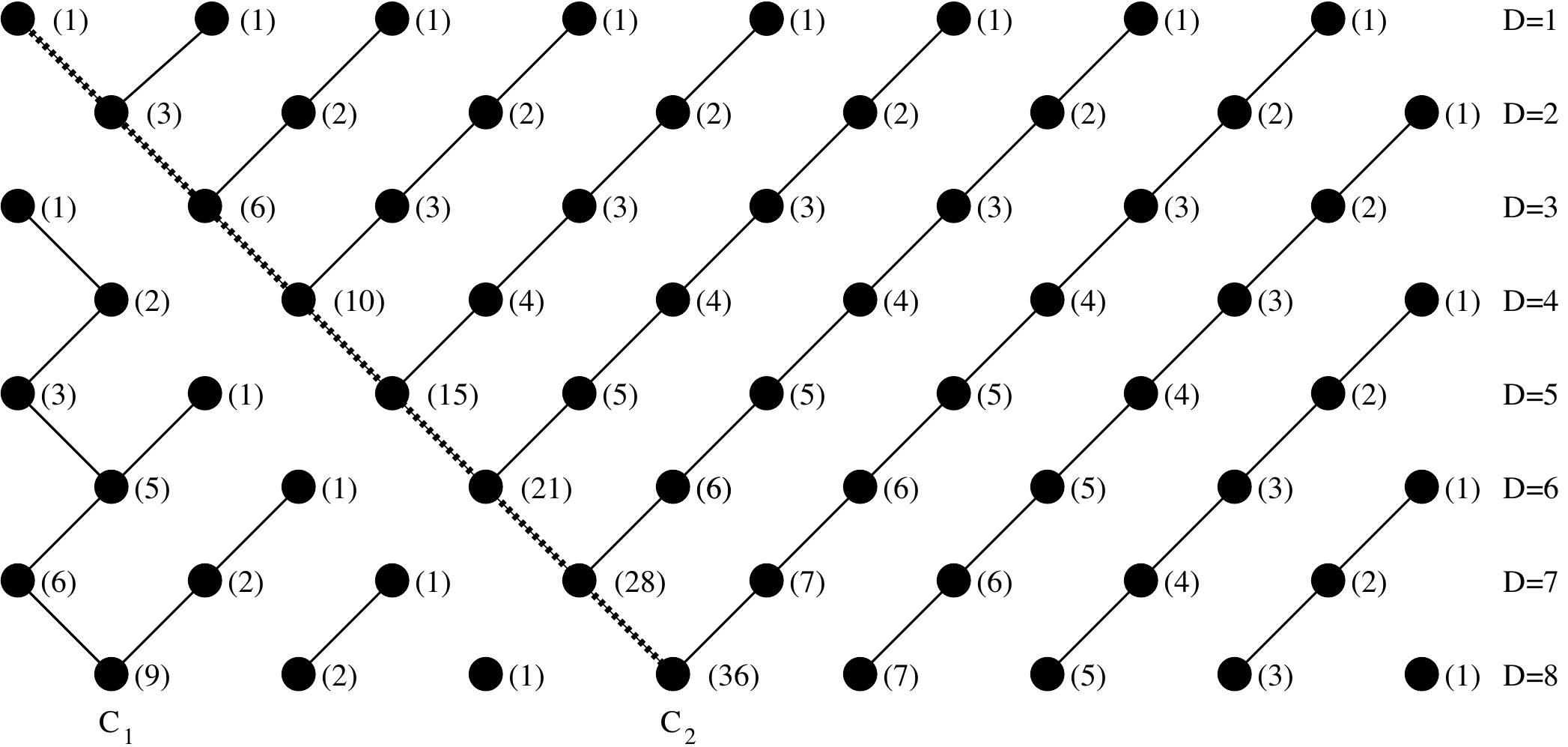}
\vspace{0.5cm}
\caption{\label{fig:vlattice} The V lattice network of $D=8$ layers with $8$ sites per layers. The number in bracket beside each site denotes its weight bearing capacity. Solid lines show connection between the sites.  The beaded line is the trunk of the V-cluster $C_2$.}
\end{center}
\end{figure}

The base model has a very specific and unique realization which incorporates the largest possible V shaped cluster that the base model could have. The V-cluster includes all the sites in the topmost layer, and ($M-D+1$) sites in the $D$th layer. One of the arms of the V constitutes the trunk, and all other connections run parallel to the arm of the V that is opposite to the trunk. The V-cluster is the largest possible cluster the base model could have. The trunk of this V-cluster bears the largest possible trunk capacity of the base model. We call this special lattice the V lattice. Figure.\ref{fig:vlattice} shows the V lattice configuration. Structures similar to the V lattice can be seen in riverine deltas \cite{riverdelta}, in Martian gullies \cite{marsgully}, and in granular flows \cite{shinbrot}, if the channels of maximal flow capacity are considered\footnote{In the natural structures, the channels of high capacity can have some channels of lower capacity joining them, leading to an overall symmetry between left and right connections, however one direction is favored by the channels of high capacity, due to some feature like the nature of the geographic terrain, leading to variants of the V structure for the high capacity channels alone.}.

\section{The V lattice: The critical case for transport}
\label{sec:2}

In this section we will discuss the critical nature of the V lattice, a special realization of the base lattice. The V lattice exhibits critical behavior for avalanche phenomena \cite{pramana}, site percolation phenomena \cite{pre12}, and capacity distribution and failure rate \cite{pre11} quite distinct from  the behavior seen  for the base lattice model, manifesting its criticality. Here, we focus only on the avalanche and percolation properties of the V lattice.

\subsection{Avalanche times distribution of the V lattice}
\label{subsec:2}
The behavior of avalanche processes for a given  network topology provides an interesting example of the interplay between the nature of a transport process and the topology of the substrate. Here we aim to study the avalanche process of weight transmission on the V lattice. The avalanche is defined  here in terms of weight transmission on the network.   

To initiate the avalanche process on the network, some  test weight $W_{test}$ is deposited on a randomly chosen site in the first layer of the network. Since our base network is a directed network, the flow of weight transmission takes place in the downward direction. The site in the first layer absorbs weight equal to its capacity and transmits the excess weight to its neighboring site it is connected to in the layer below. This process of weight transmission continues till there is no excess weight left, and the transmission is successful. If the $W_{test}$ is sufficiently  large that it  reaches the last layer without being fully absorbed, which completes one cycle of weight transmission, the excess weight is then deposited randomly on a site in the first layer and the second cycle of weight transmission starts here. If in the next cycle, the receiving site happens to be the one which has already saturated its capacity in the previous cycle, and now it does not have spare capacity to absorb any weight, the transmission is then said to be failed at that site. Otherwise the weight transmission continues till all excess weight is absorbed, and the transmission is said to be successful in this case. This process of the cycling of weight transmission through the lattice network is defined as an avalanche. The duration of an avalanche, or avalanche time, is defined to be the total number of layers traversed during all cycles of successful weight transmission by the test weight in the network.

We discuss the distribution of avalanche times of the V lattice. The avalanche times distribution $P(t)$ is, in fact, the distribution of the number of layers traversed during all cycles of successful avalanche transmissions by a test weight placed at a random site in the first layer for any lattice. 

\begin{figure}[htb]
\begin{center}
\begin{tabular}{cccc}
(a)&
\hspace{-0.5cm}
\includegraphics[height=4.5cm,width=5.5cm]{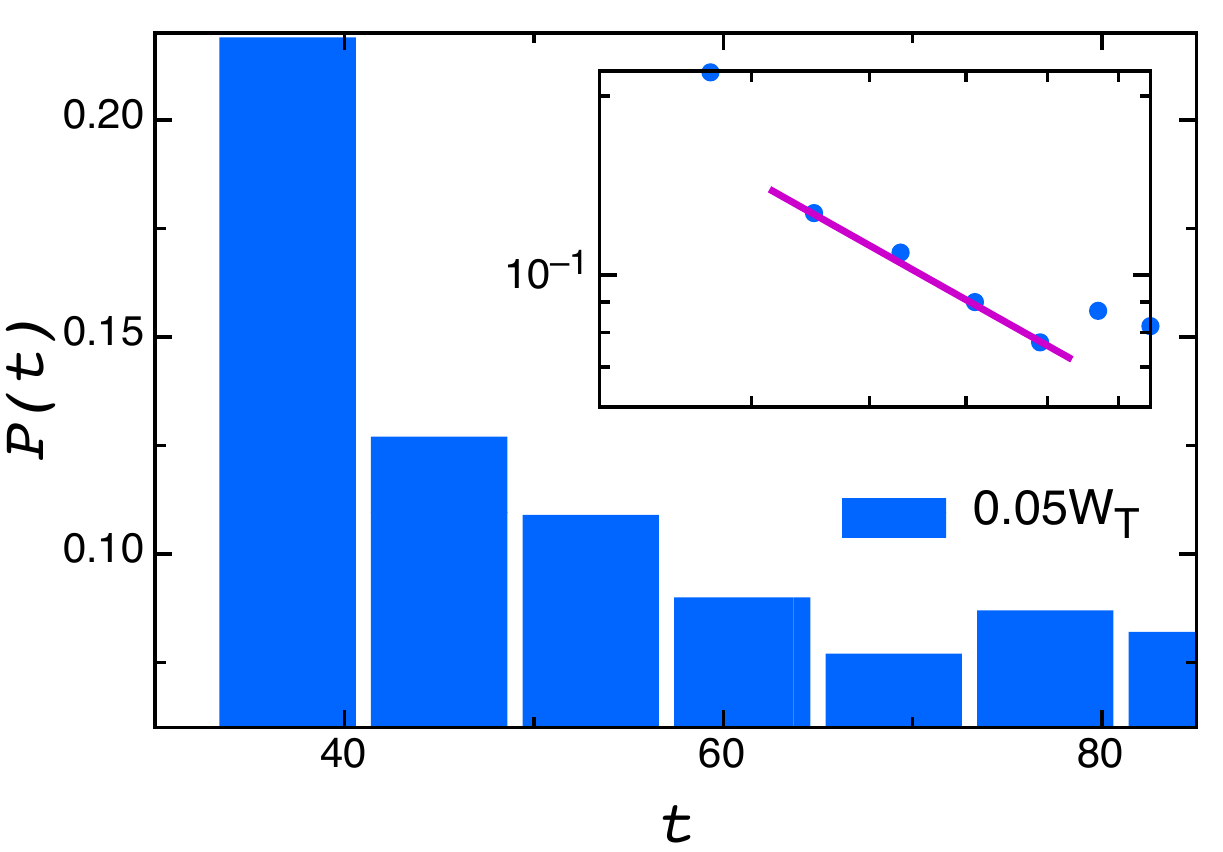}&
(b)&
\hspace{-0.5cm}
\includegraphics[height=4.5cm,width=5.5cm]{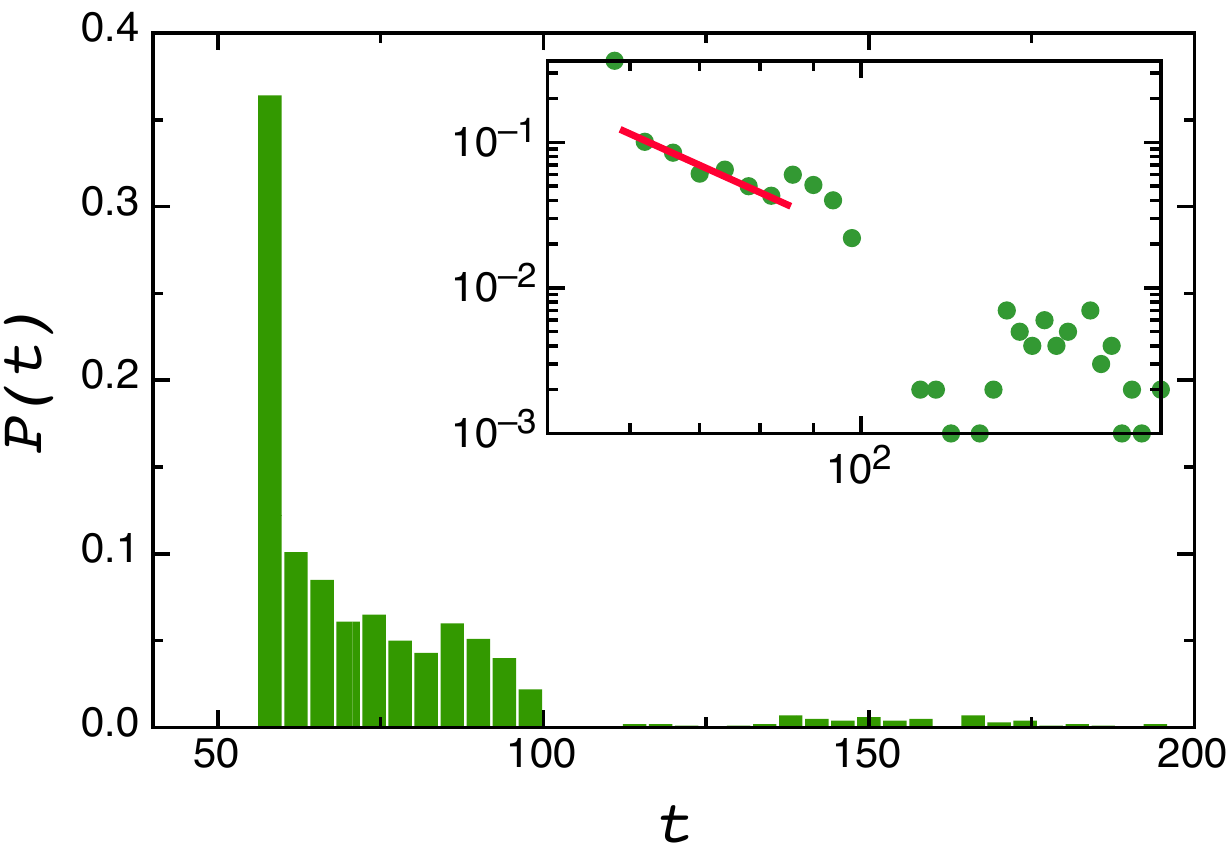}\\
\end{tabular}{}
\caption{The avalanche time distribution $P(t)$ corresponding to $1000$ realizations for the V lattice network of side $M=100$ when tested for weights equal to (a) $0.05W_T$, and (b) $0.2W_T$. Small regimes for $0.05W_T$ ( as shown in inset of (a) ) and $0.2W_T$ ( as shown in inset of (b) ) display power law \mbox{$i.e.$} $P(t)\sim t^{-\alpha}$ behavior with exponent $\alpha=1.18$ and $\chi^2=0.0009$, and $\alpha=2.96$ and $\chi^2=52.197$ respectively.}
\label{fig:vlat}
\end{center}
\end{figure}

\begin{figure}[t]
\sidecaption[t]
\includegraphics[scale=0.55]{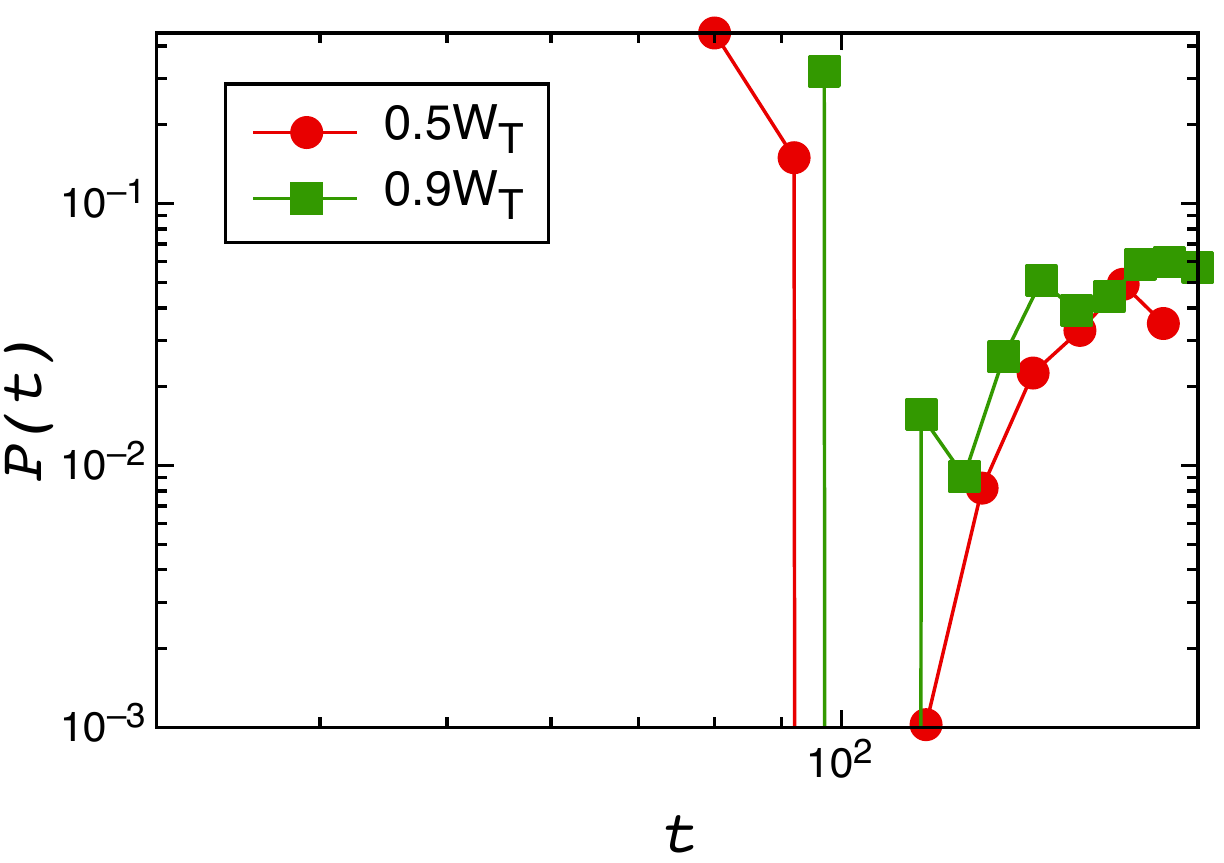}
\caption{The avalanche times distribution $P(t)$, corresponding to $1000$ realizations, for the V lattice network of side $M=100$ when tested for weights equal to (a) $0.5W_T$, and (b) $0.9W_T$. No power law regime is seen in the distributions corresponding to $0.5W_T$, and $0.9W_T$.}
\label{fig:vlat_59wt}
\end{figure}

\begin{figure}[t]
\sidecaption[t]
\includegraphics[scale=.55]{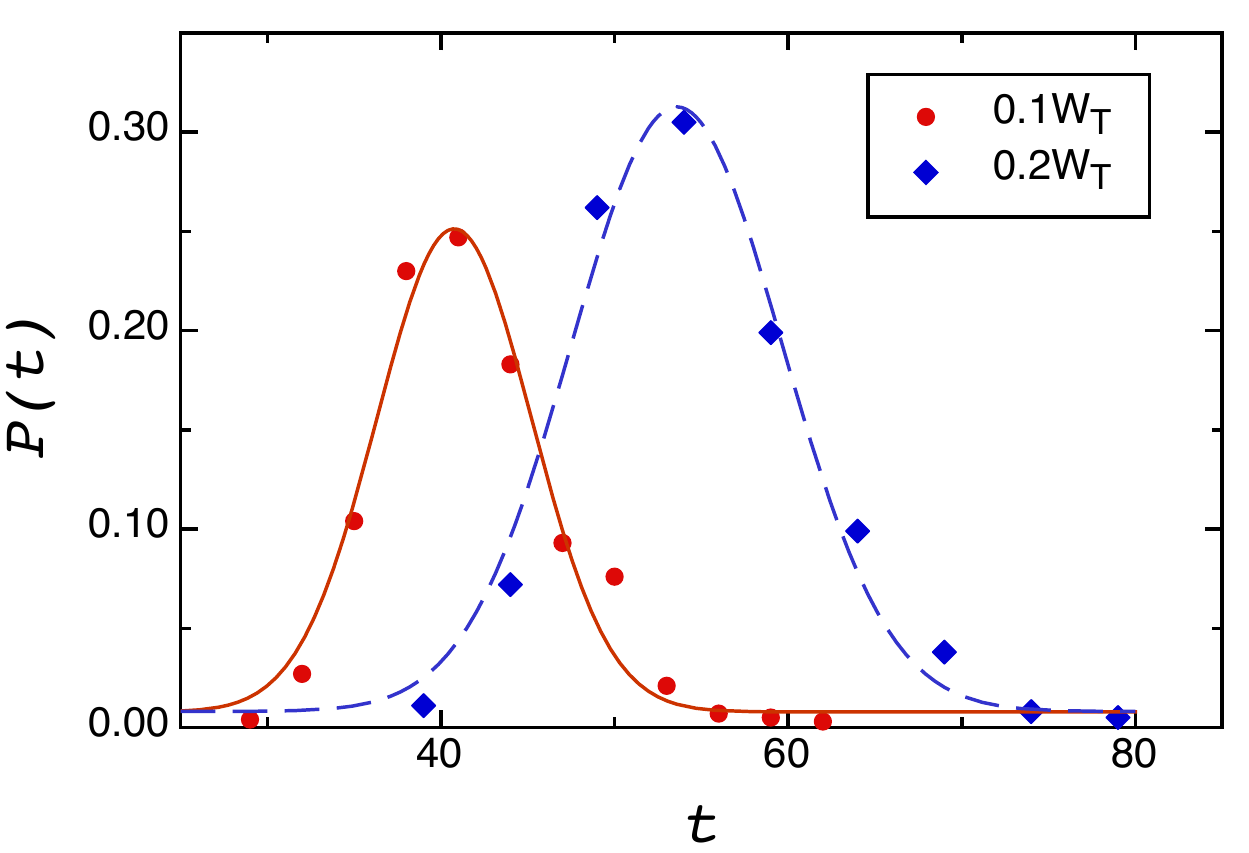}
\caption{The avalanche times distributions $P(t)$ for the base lattice, corresponding to $1000$ realizations of networks of side $M=100$, exhibits Gaussian behavior when tested for weights equal to $0.1W_T$  and $0.2W_T$, where $W_T$ is the trunk capacity of base lattice.}
\label{fig:base_aval}
\end{figure}

The avalanche times distribution $P(t)$ of the V lattice has been studied by Kachhvah and Gupte \cite{pramana} for test weights which are fractions of the trunk capacity of the V lattice. It has been observed that for this lattice, the avalanche times distribution $P(t)$ displays a power law regime, i.e., $P(t)\sim t^{-\alpha}$, which gradually starts disappearing as the test weight starts approaching the trunk capacity of the V lattice.
Figs. \ref{fig:vlat}, and \ref{fig:vlat_59wt} display the avalanche times distributions, corresponding to $1000$ realizations, for the V lattice. The power law distribution is seen in the V lattice as weight transmissions on the V lattice can achieve success at any one of the layers. This behavior of the existence and subsequent disappearance of the power law regime in the distribution has been one of the indications that the V lattice is a critical case of the base lattice. It is to be noted that the avalanche times distribution averaged over many realizations of the base lattice shows Gaussian behavior Fig. \ref{fig:base_aval} \cite{pramana}, and is quite distinct from the power-law behavior seen here.

\subsection{Explosive percolation on the V lattice}
\label{subsec:2}
The problem of site percolation has frequently been studied on networks due to its relevance for the problem of information transfer on the network.
Therefore, it is important to study site percolation on the hierarchical networks. The site percolation problem is set up as follows:
Typically, the transmission of information is modeled by packets of information hopping on the sites of the substrate networks. These packets could be the data packets of information in the Internet, which links computers of heterogeneous and high capacities capable of transmitting packets at high rates. For our hierarchical networks, the packets are deposited at a randomly chosen site on the topmost layer of the network. Each site retains the number of packets which saturates its capacity and the remaining packets are transmitted further. A packet at a given site sees the nearest neighbor sites linked to itself. If the targeted neighboring site is not fully occupied (i.e., it has not saturated its capacity), the packet moves there and looks for the next site that has spare capacity. If the target site is fully occupied, then the packet stops on the site which it occupies, and the transmission of the packet ends at that site. In this fashion, all packets hop from one site to another according to the vacancy available on the neighboring sites, and this process continues till all the packets come to rest at some site.
When the packet transmission has come to rest, at that time some of the sites would be occupied while the other would remain unoccupied or free. In this scenario, the network is composed of two sub-networks one of the unoccupied or free one and other of the occupied one, with the size of each network being a measure of the saturated or available capacity on the network. 
Hence, we study the transition to percolation in the occupied or unoccupied sub-networks. For the free sub-network, one anticipate a transition from the percolating to the non-percolating state when the packet density $\mu$ increases, where the packet density is the ratio of total number of packets to total number of sites in the network.
In order to analyze the transition, we simulate the stochastic dynamics of packet transmission in the V lattice for different values of the packet density $\mu$ \cite{pre12}.

To study the percolation transition, the order parameter is defined to be the percolation strength $S=S_m/L$, where $S_m$ indicates the number of sites belonging to the largest connected cluster of the unoccupied sub-network and $L$ denotes the total number of sites in the network. The complement of the percolation strength $S$ is defined as $S_1=(1-S)$, which is a measure of the fraction of occupied sites or of the size of the occupied sub-network. We note that $S$, the size of the connected cluster of unoccupied sites is a measure of the capacity available for transport for a given packet density. This goes to zero at the jamming transition where no more capacity is available, and its complement $S_1$ goes to the size of the lattice. Here $S_1$ is compared with the size of the percolating cluster which spans the size of the lattice at the percolation transition.

\begin{figure}[htb]
\begin{center}
\begin{tabular}{cccc}
(a)&
\hspace{-0.6cm}
\includegraphics[height=4.5cm,width=5.5cm]{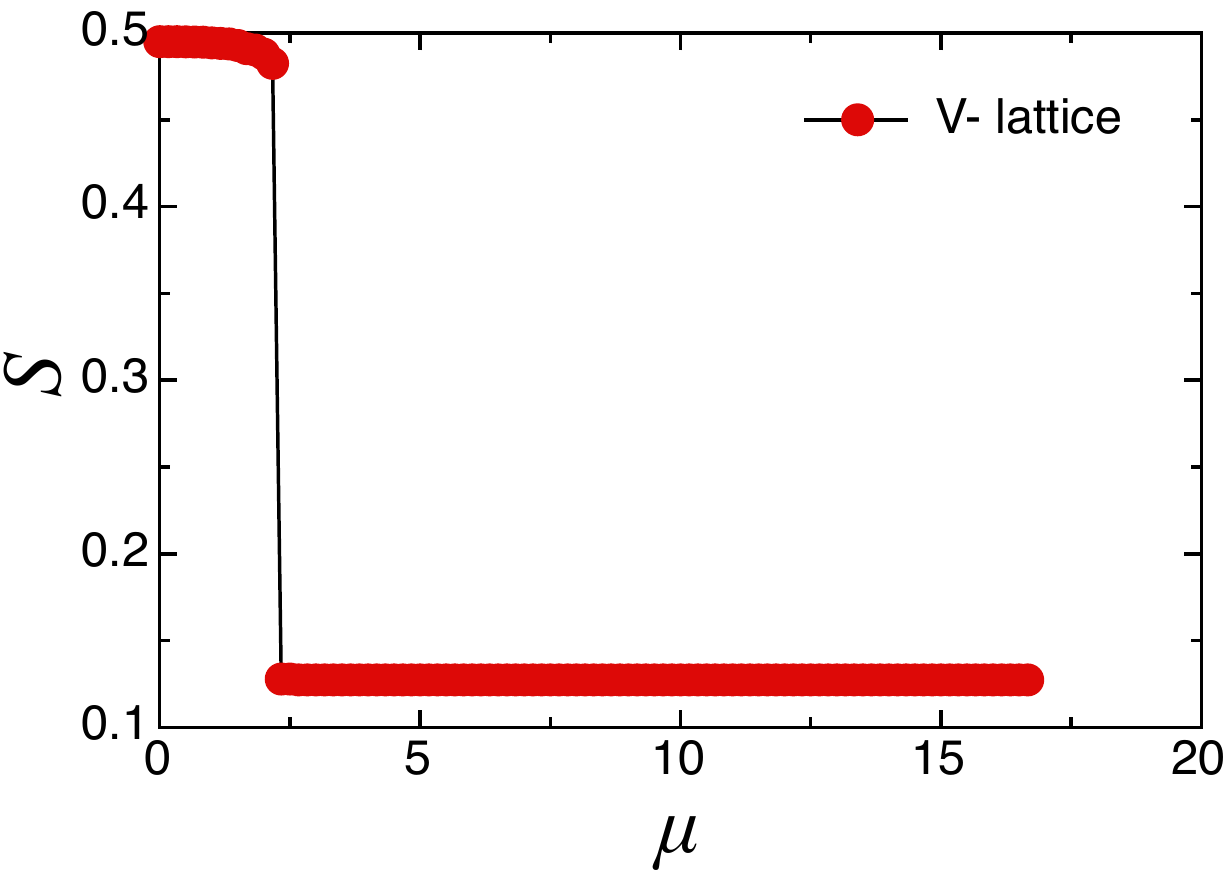}&
(b)&
\hspace{-0.6cm}
\includegraphics[height=4.5cm,width=5.5cm]{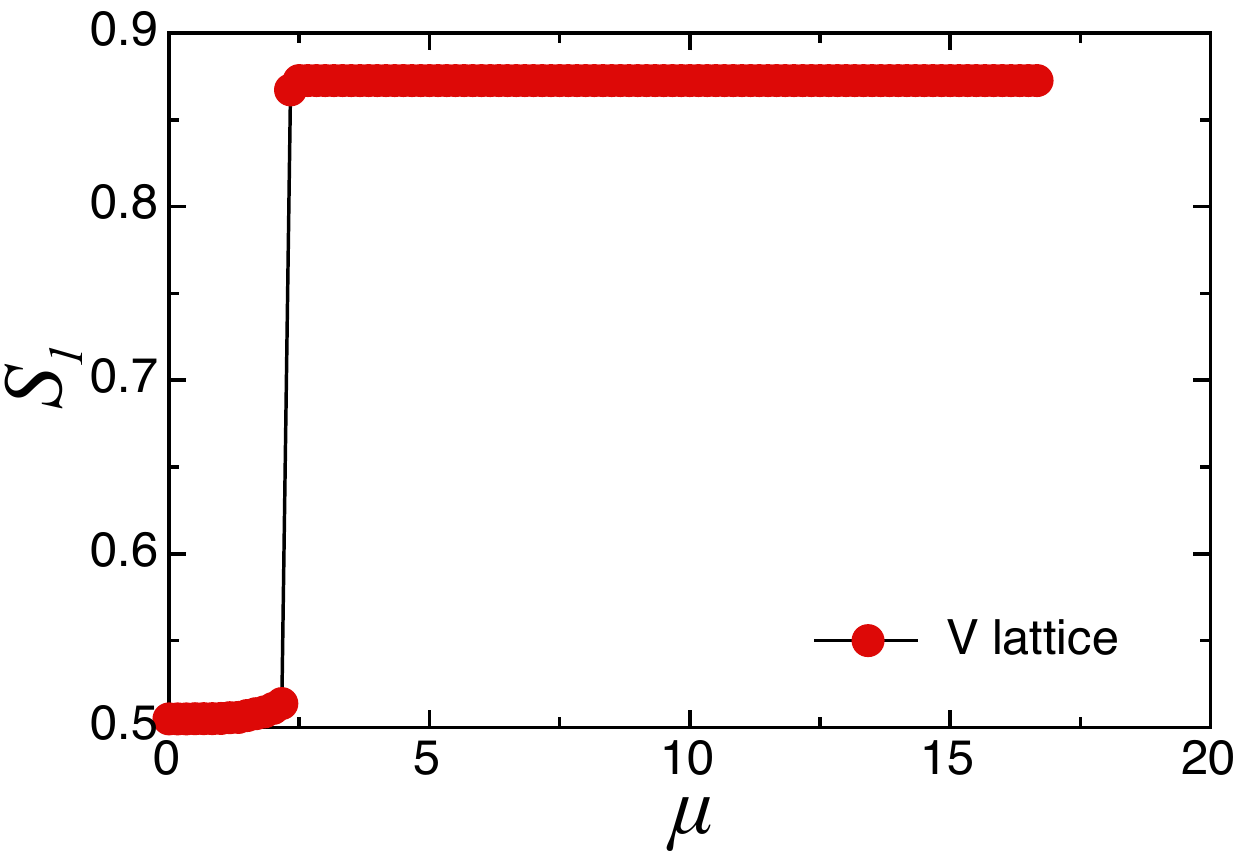}\\
\end{tabular}{}
\caption{\label{fig:perco} (a) The order parameter $S$ and (b) its complement $S_1$ as a function of packet density $\mu$, averaged over $500$ realizations, for the V lattice networks of $(100\times100)$ sites.}
\end{center}
\end{figure}

\begin{figure}[t]
\sidecaption[t]
\includegraphics[height=4.5cm,width=6.5cm]{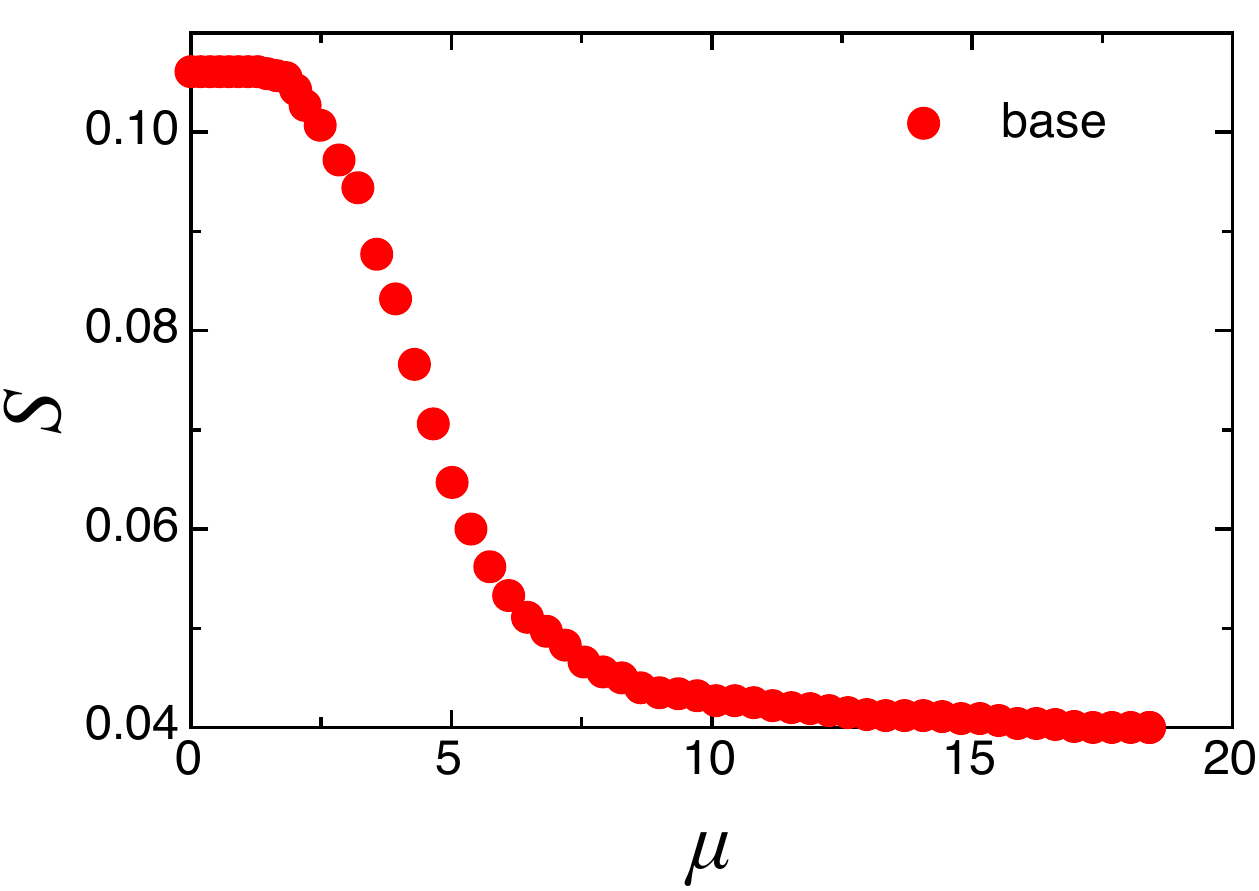}
\caption{The order parameter $S$ as a function of packet density $\mu$, averaged over $500$ realizations, for the free sub-network in the base lattice network of $(100\times100)$ sites.}
\label{fig:base_percoS} 
\end{figure}

The numerical study by \cite{pre12} shows that the percolation transition in the V lattice is a discontinuous or explosive one, which can be observed from the Figure. \ref{fig:perco}, and laws of the finite size scaling for continuous percolation transition does not hold for the V lattice. The percolation transition in the V lattice is contrary to second order continuous transition seen in the base lattice Fig. \ref{fig:base_percoS} \cite{pre12}, with associated second order percolation exponents. Here, again the V lattice shows critical behavior.

\section{Perturbed  V lattices: annihilation of criticality}
\label{sec:3}
We have seen that the V lattice is the critical case of the base model as it displays behaviors for the avalanche times distribution and site percolation quite distinct from  those seen for the base lattice. Here we are interested in exploring whether a slight perturbation introduced in the structure of the V lattice destroys the critical nature of the V lattice for avalanche and site percolation phenomena. 
For this, perturbed V lattices are generated by switching the direction of the connections running parallel to arm opposite to the trunk in the V-cluster, with some probability $0<p<0.5$, but leaving the trunk or backbone untouched.

\subsection{Avalanche time distributions for the perturbed V lattice}
\label{subsec:3}
\begin{figure}[t]
\sidecaption[t]
\includegraphics[height=5.0cm,width=7.0cm]{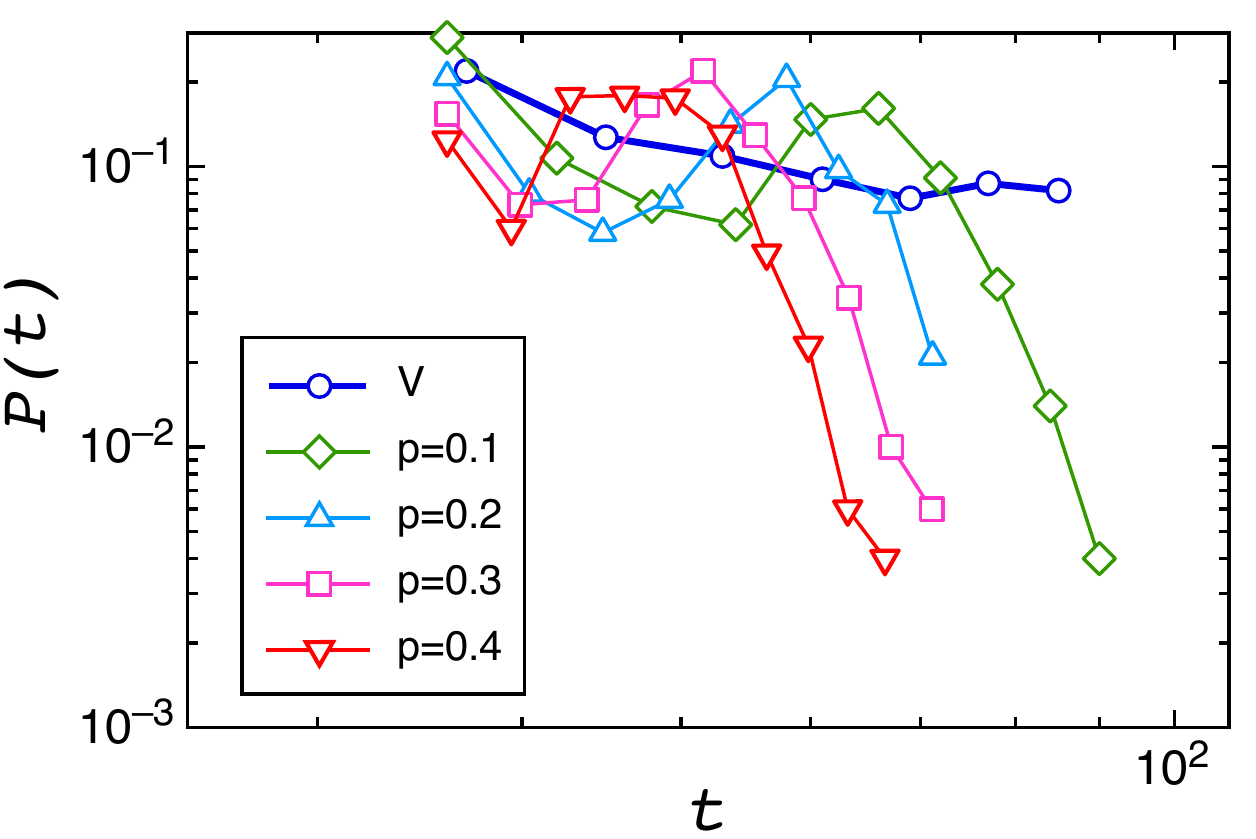}
\caption{The distributions of avalanche times of the V lattice and the perturbed V lattices obtained by switching connections with probability $p=0.1,\ldotp\ldotp\ldotp0.4$, corresponding to $1000$ realizations of a network of side $M=100$ when tested for weights equal to $0.05W_T$.}
\label{fig:vlat_perturbed}
\end{figure}

To demonstrate that the V lattice is the only realization of the base lattice which displays power law behavior for the avalanche time distribution, we studied these distributions for the perturbed V lattices, for 1000 realizations and test weights equal to $0.05W_T$ as shown in Fig. \ref{fig:vlat_perturbed}. It is observed that even a small perturbation ($p=0.1$) in the topology of the V-cluster of the V lattice, destroys the power law behavior seen for the V lattice. Thus, the V lattice is found to be the only case of the base lattice which displays power law behavior.

\subsection{The percolation transition for the perturbed V lattice}
\label{subsec:3}
\begin {figure}[htb]
\begin{center}
\begin{tabular}{cccc}
(a)&
\hspace{-0.5cm}
\includegraphics[height=4.5cm,width=5.5cm]{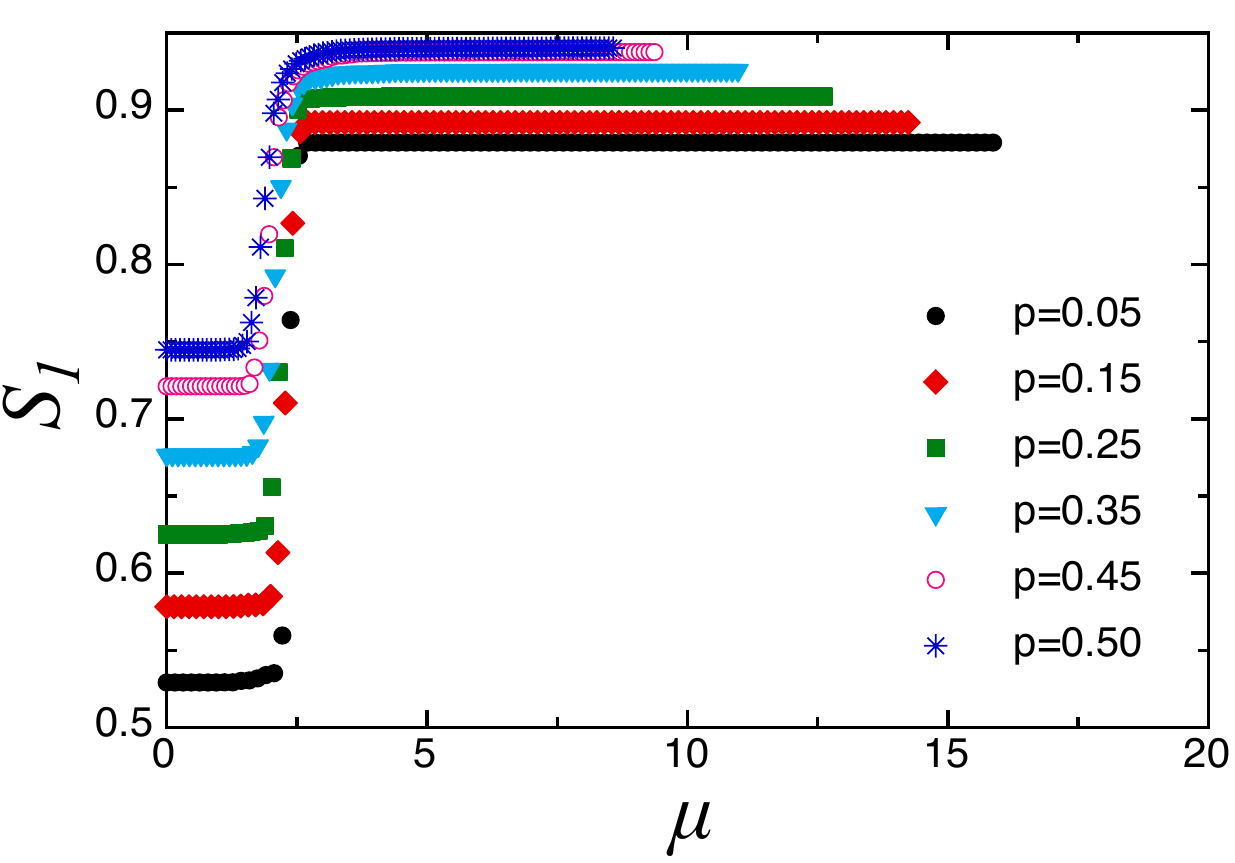}&
(b)&
\hspace{-0.5cm}
\includegraphics[height=4.5cm,width=5.5cm]{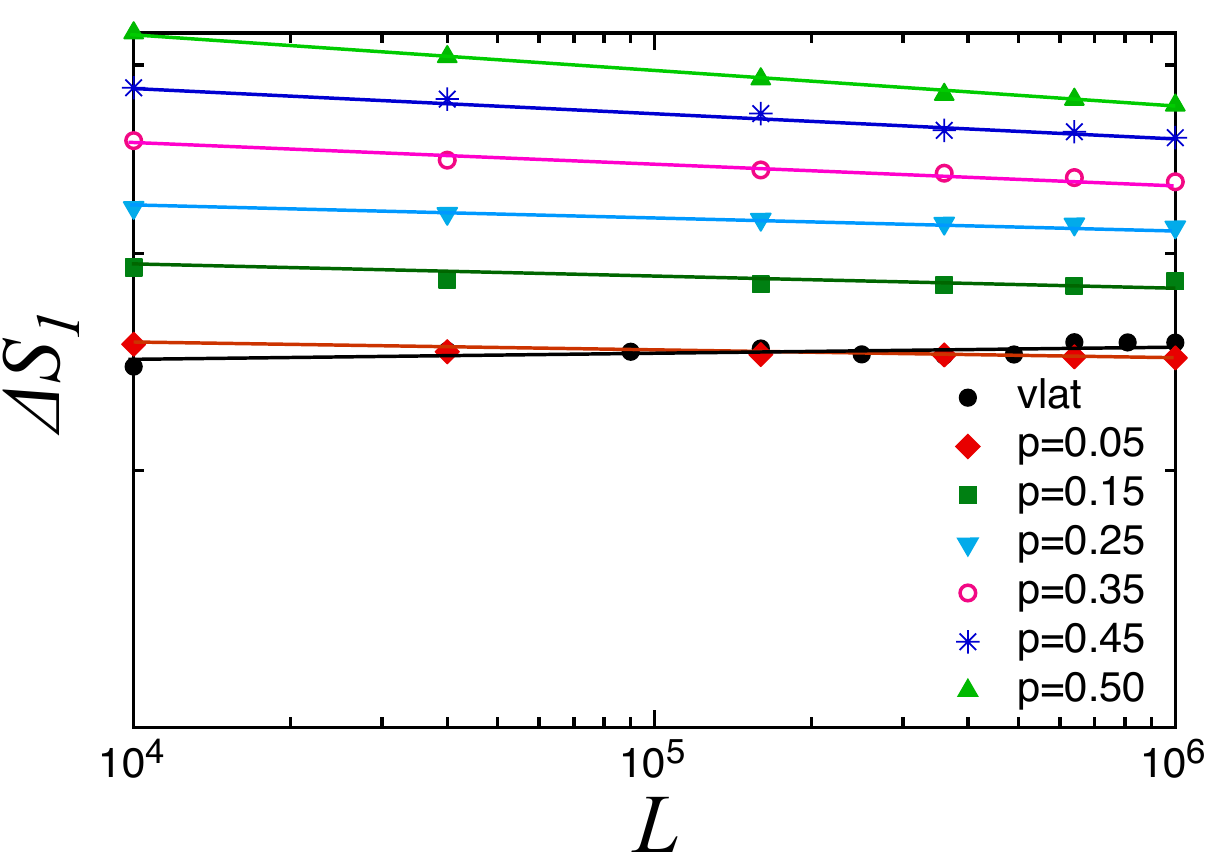}\\
\end{tabular}{}
\caption{\label{fig:vlat_jump} Plot of (a) the order parameter $S_1$ of the occupied sub-lattice as a function of critical packet density $\mu$ for different values of $0<p<0.5$. (b) The largest jump size $\Delta S_1$ of the V lattice, scales as a power law as function of the lattice size $L$ \mbox{$i.e.$}, $\Delta S_1 \sim L^{-\phi}$, where $\phi$ is $0.0019$, $0.0023$, $0.0029$, $0.0059$, $0.0079$ and $0.011$ for $p$ equal to $0.05$, $0.15$, $0.25$, $0.35$, $0.45$, $0.50$, respectively.}
\end{center}
\end{figure}

Again, in order to demonstrate that the V lattice is the only realization of the base lattice which displays explosive percolation transition, we explore the percolation transition on the perturbed V lattices. The order parameter $S_1$ is computed as a function of the packet density $\mu$ (see Fig. \ref{fig:vlat_jump} (a)) for the V lattice perturbed by different values of $p$, \mbox{$i.e.$}, $0<p<0.5$. From Fig. \ref{fig:vlat_jump}(a), one can notice that as the perturbation in the V lattice is increased (\mbox{$i.e.$}, as $p$ increases), the size of the jump $\Delta S_1$ also reduces, where $\Delta S_1$ is the difference between the values of $S_1$ before and after the transition. In Fig. \ref{fig:vlat_jump}(b), the size of the largest jump $\Delta S_1$ is plotted, corresponding to different $p$, for different lattice sizes $L$. It is apparent from Fig. \ref{fig:vlat_jump}(b) that the largest jump size $\Delta S_1$, scales as a power law with system size $L$, defined as:
\begin{equation}
\Delta S_1 \sim L^{-\phi}.
\end{equation}
The above relation is similar to the relation between the largest jump size and the system size in \cite{nagler} to show the discontinuity of the percolation transition. We find that $\phi>0$ for all processes of the V lattice corresponding to different $p$. However, the values of $\phi$ are quite small. If,  in the thermodynamic limit of infinite system size, we have 
\begin{equation}
lim_{L\rightarrow \infty}\Delta S_1=0.
\end{equation}
i.e. if, in the limit of infinite system size, the size of the largest jump $\Delta S_1$ for the giant cluster goes to zero, then transitions are said to be weakly discontinuous.
The percolation transitions in the perturbed V lattices are weakly discontinuous. We conclude that even a slight perturbation introduced in the V lattice destroys the nature of the percolation transition in the V lattice.

We thus conclude that the V lattice is indeed a critical case of the base lattice, as  even a slight distortion in the structure of the V lattice destroys its critical nature. 

\section{Conclusions}

To summarize, we have observed that in the case of branching hierarchical structure, the lattice with V-shaped clusters shows special behavior for transport processes which use this lattice as the substrate. Avalanche processes on this lattice show power law behavior, and percolation behavior 
on this lattice belongs to the explosive percolation class.  This behavior sharply contrasts with the behavior seen for transport on typical realizations of the hierarchical networks, where avalanche transmissions are Gaussian, and the transition to percolation is of the usual second order type. Small perturbations to the V-cluster, rapidly destroy the special behavior, indicating the critical nature of the lattice. We note that this is one of the few examples where the nature of the substrate topology has led to the identification of a transition of the explosive percolation class. We hope our results go some way towards the identification of special topologies where critical behavior is observed for transport properties. Such an identification maybe of utility in practical systems like power grids and computer and communication networks.

%%%%%%%%%%%%%%%%%%%%%%%% referenc.tex %%%%%%%%%%%%%%%%%%%%%
% sample references
% 
% Use this file as a template for your own input.
%
%%%%%%%%%%%%%%%%%%%%%%%% Springer%%%%%%%%%%%%%%%%%%%%%%%%%%
%
% BibTeX users please use
% \bibliographystyle{}
% \bibliography{}
%

\end{document}